# Hypermultiplexed off-chip hologram by on-chip integrated metasurface

*Xianjin Liu, Zhanying Ma, Dasen Zhang, Qiwen Bao, Zhenzhen Liu, and Jun-Jun Xiao\**

**\*Corresponding author: Jun-Jun Xiao,**
College of Electronic and Information Engineering, Harbin Institute of Technology (Shenzhen), Shenzhen 518055, China; Shenzhen Engineering Laboratory of Aerospace Detection and Imaging, Harbin Institute of Technology (Shenzhen), Shenzhen 518055, China; and College of Integrated Circuits, Harbin Institute of Technology (Shenzhen), Shenzhen 518055, China,
E-mail: eiexiao@hit.edu.cn

Xianjin Liu, Zhanying Ma, Dasen Zhang, Qiwen Bao, Zhenzhen Liu,
College of Electronic and Information Engineering, Harbin Institute of Technology (Shenzhen), Shenzhen 518055, China; and Shenzhen Engineering Laboratory of Aerospace Detection and Imaging, Harbin Institute of Technology (Shenzhen), Shenzhen 518055, China

Zhenzhen Liu,
Research Center for Advanced Optics and Photoelectronics, Department of Physics, College of Science, Shantou University, Shantou 515063, China

**Abstract:** The waveguide-integrated metasurface introduces a novel photonic chip capable of converting guided modes into free-space light. This enables functions such as off-chip beam focusing, steering, and imaging. The challenge lies in achieving hyper-multiplexing across diverse parameters, including guided-wave mode type, direction, polarization, and notably, multiple wavelengths. Here, we introduce a comprehensive end-to-end inverse design framework, rooted in a physical model, for the multifunctional design of on-chip metasurfaces. This framework allows for metasurface optimization through a target-field-driven iteration process. We demonstrate a hypermultiplexed on-chip metasurface capable of generating red-green-blue holograms at multiple target planes, with both independent and cooperative control over guided-wave direction. Significantly, the proposed method streamlines the design process utilizing only the positions of meta-atoms as the design variable. We demonstrate 9 independent holographic channels through a combination of wavelength and distance multiplexing. Moreover, by incorporating the excitation direction into the design, the metasurface produces a total of 36 distinct holograms. The robustness of these results against fabrication discrepancies is validated through 3D full-wave electromagnetic simulations, aligning well with advanced manufacturing techniques. Our research presents a universal design framework for the development of multifunctional on-chip metasurfaces, opening up new avenues for a wide range of applications.

## 1. Introduction

Metasurfaces have emerged as a promising platform for manipulating light at the subwavelength scale, offering unprecedented control over the propagation, amplitude, polarization, and phase of electromagnetic waves.[1–3] A key advantage of metasurfaces is their capacity to simultaneously customize various properties of light and operate in a multiplexed manner. This is achieved by carefully designing the geometry, composition, and arrangement of subwavelength elements, known as meta-atoms. By judiciously constructing metasurface structures, it becomes feasible to manipulate electromagnetic wave characteristics independently or cooperatively such as amplitude, phase, polarization, spectral, and spatial dispersion. These artificially engineered structures have revolutionized a wide range of applications in optics and nanophotonics, including imaging, sensing, display technology, holography, optical computation, diffractive optical neural networks, and optical communication systems.[4–8]

Despite the rapid advancements in metasurfaces for applications such as flat optics in free-space,[9,10] free-space to chip coupling and conversion,[11–13] and on-chip or surface-bound wave manipulation,[14–17] there is a burgeoning interest in managing emitted waves off-chip with meticulously engineered metasurfaces.[18,19] These guided-wave-driven metasurfaces broaden the horizons of meta-optics and have recently achieved significant progress.[20] Notable examples include their deployment in launching on-chip signals into fibers[21] and the arbitrary aperture synthesis via microwave metasurface antennas.[22] Within this off-chip emission metasurface paradigm, artificial micro- and nano-structures are crafted to interface with waveguide modes (or surface-bound modes) and free-space optical fields, thereby enabling precise control over the diffraction of guided modes into the ambient environment.[23]

The integration of guided-wave-driven metasurfaces presents several advantages over light-beam-driven metasurfaces. Firstly, it enables effective coupling and precise manipulation of light within a guided mode, thereby enhancing light-matter interactions and potentially improving device performance. Secondly, it enables the development of compact, on-chip optical devices, resulting in a smaller footprint and better integration with other system components.[14,20,24–29] Thirdly, the design flexibility of metasurfaces permits the implementation of sophisticated functionalities, such as beam steering, focusing, polarization conversion, and spectral filtering. These functionalities, which are extensively explored for free-space optics, can theoretically be seamlessly adapted to the off-chip version within a compact device footprint. Moreover, the off-chip emitted beams are uniquely free from zero-order background interference, an issue that is inherent in free-space to free-space metasurface applications.

Designing metasurfaces for operation across multiple channels, particularly regarding spectral response, has posed a significant challenge.[3,30] This difficulty stems from the reliance on forward-design approaches, which typically involve searching for a phase modulation profile tailored to a single spectral channel, that is, operation at a specific wavelength. Complex objectives are often broken down into sub-objectives, with phase modulation maps for each sub-objective obtained using the Gerchberg-Saxton (GS) algorithm.[3] These phase maps are then encoded onto different optical channels. For instance, wavelength-encoded channels can be utilized to achieve RGB color holography.[4, 31] Additionally, polarization multiplexing holography, which relies on polarization-encoded channels, has been realized.[32] It is even feasible to encode holographic information into orbital angular momentum (OAM) mode channels, facilitating ultra-high-density multiplexing holography.[33–37] Generally, these methods require meticulously designed meta-atoms with an increased number of geometric degrees of freedom to enhance the number of encoding channels and minimize crosstalk. However, these design strategies for free-space metasurface multiplexing are not straightforwardly transferable to on-chip metasurfaces. This is due to the sequential nature of guided-wave driving, which contrasts with the parallel operation on digitized meta-atoms typical of free-space applications.

However, there are also various successful design methods for on-chip metasurface functional multiplexing. For example, different OAM modes and holograms can be emitted off-chip by switching the guided-wave excitation direction.[18] Additionally, introducing diffraction distance multiplexing can enable more holographic channels.[38,39] Nevertheless, most methods are targeted at single-wavelength applications, despite significant progress. Recently, RGB color holograms have been achieved by encoding the three-wavelength phase maps into two orthogonal detour and Pancharatnam-Berry (PB) phase channels, respectively.[40,41] This strategy demands more geometric degrees of freedom for the meta-atoms, complicating the design and fabrication of metasurfaces. Furthermore, the limited number of reconstructed RGB color holograms constrains the practical applications of on-chip metasurfaces. Therefore, developing on-chip metasurfaces for high-capacity RGB color holography is extremely challenging.

In fact, the end-to-end optimization approach has proven successful in designing metasurfaces,[30,42] as it allows for the creation of a differentiable model that connects target physical quantities with the metasurface geometry. Utilizing backpropagation schemes such as the gradient descent algorithm, one can iteratively update the geometry to guide the system toward an optimal solution. In this work, we propose and demonstrate an approach for the inverse design of guided-wave-driven metasurfaces for multi-color and multi-plane holographic image projection (see **Figure 1**). Specifically, we present an on-chip metasurface with a single-cell (meta-atom) capable of projecting pre-designed RGB images across various planes at different off-chip distances. Moreover, our method can increase the number of holographic channels by merely adjusting the objective function, without expanding the geometric degrees of freedom required for encoding various prior phases. We can even double the holographic channels corresponding to the metasurface for guided wave excitation in both the $+\hat{x}$ and $-\hat{x}$ directions by simply manipulating the loss function. This enables us to achieve two distinct sets of RGB color holographic images in the opposite directions of the excitation wave, totaling 18 holograms. Similarly, the $+\hat{y}$ and $-\hat{y}$ directions can also generate the same number of holographic channels without obviously affecting the $\pm\hat{x}$ channels. Thus, we are able to achieve up to 36 holographic images with a single metasurface. Additionally, we introduce a co-design strategy that maps the displacement of meta-atoms in the orthogonal and $\hat{y}$ directions to generate the same target field, effectively enhancing the hologram's peak signal-to-noise ratio (PSNR). Consequently, we provide a versatile framework for the inverse design of multifunctional on-chip metasurfaces, which significantly simplifies the design process, reduces metasurface complexity, and enhances multiplexed functionality.

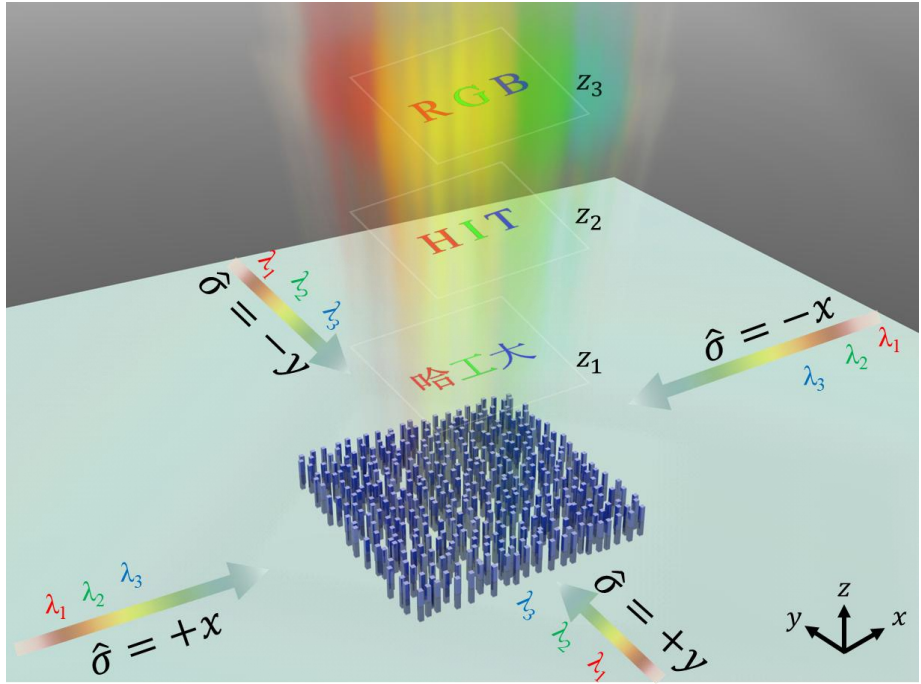

**Figure 1.** Schematic of the on-chip metasurface for multiplexed optical holographic images containing

three wavelengths $\lambda$, three reconstruction distances $z$, and four guided wave incoming directions $\hat{\sigma}$.

## 2. Physical Model for Off-Chip Scattering and the End-to-End Inverse Design Framework

Let us begin by establishing the physical model for the guided-wave-driven metasurface shown in Figure 1. Unlike metasurfaces that directly modulate free-space wavefronts directly, on-chip metasurfaces engage with the waveguide mode field. They extract and scatter a fraction of its energy into the free space. Consequently, the phase of the emitted wave results from a combination of the abrupt phase inherent to the meta-atom structure and the phase of the guided wave mode at the point of interaction. This combined phase is referred to as the detour phase.[37] For the sake of simplicity, we initially focus on the scenario where the guided wave propagates solely along the $\hat{\sigma} =+ \hat{x}$ direction. The phase of the extracted wave at a given coordinate $x$ is determined by

$$\varphi_{+\hat{x}}(x, y) = \varphi_0 + \varphi_a(x, y) + \beta(\lambda)x, \qquad (1)$$

where $\varphi_0$ represents the initial phase of the incident wave, $\varphi_a(x, y)$ denotes the abrupt phase induced by the meta-atom scattering.[28,43] This phase change is a consequence of the inherent light conversion process, which can manifest as changes in the propagation phase, resonance phase, or PB phase, etc. In Eq. (1) the term $\beta(\lambda) = 2\pi x n_{eff}/\lambda$ determines the detour phase of the waveguide mode [see **Figure 2**(a)], while $\beta(\lambda)$ represents the propagation constant of the waveguide mode, and $n_{eff}$ is the effective refractive index of the fundamental mode (e.g., TE$_0$).

For multi-wavelength applications, the metasurface generates varied wavefronts for different wavelength, determined by the effective propagation constant, as shown in Figure 2(b). Note that the conversion efficiency of the same meta-atom would be slightly dependent on the wavelength, as well as the position due to the guided-wave energy decay [28]. However, we neglect the scattering amplitude variation considering the relatively small volume of the meta-atom and the small area of the metasurface.

As shown in Figure 2(c), the metasurface is designed comprising $M \times N$ square pixels of length $\Lambda$, spanning total area of $M\Lambda \times N\Lambda$. All meta-atoms are assumed to be identical, so we can neglect $\varphi_a$ in Eq. (1). Therefore, the scattering phase of the meta-atom at the metasurface pixel $(i, j)$ is determined by

$$\varphi_{+\hat{x}}(i, j) = \varphi_0 + \beta(\lambda)[\Lambda(i - 1) + \Delta x], \qquad (2)$$

where $\Delta x$ denotes the relative displacement of the meta-atom in the pixel $(i, j)$ within a regular grid with cell size $\Lambda \times \Lambda$ [see the zoom-in inset of Figure 2(c)]. Therefore, the desired phase map can be achieved by adjusting $\Delta x \subset (-\Lambda/2, \Lambda/2)$ but possible overlapping between neighboring meta-atoms is avoided by setting a constraint on the neighboring metaatom distance $|x_{(i,j)} - x_{(i+1,j)}| \geq \Delta_{min}$ In the multi-wavelength scenarios, selecting an appropriate value for $\Lambda$ ensures that $\Lambda\beta(\lambda) \geq 2\pi$ could be satisfied for all operating wavelengths (see Figure S2 in Supporting Information), enabling the metasurface to achieve complete wavefront control. For the guided wave propagation towards $-x$ direction, the scattering phase of the meta-atom at the pixel $(i,j)$ is expressed as

$$\varphi_{-\hat{x}}(i,j) = \varphi_0' + \beta(\lambda)(\Lambda(M - i + 1) - \Delta x). \tag{3}$$

Similarly, when the guided wave propagates along the $\pm y$ directions, we can encode the phase map into $\Delta y$. For demonstration and verification purpose, we have constructed a one-dimensional on-chip metalens by arranging 20 metatoms on a strip waveguide, which extracts and focuses the light field towards the target point above the chip (see Supporting Information, Note S1 for details).

Creating an on-chip metasurface for wavelength multiplexing with the forward design method is challenging, using solely $\Delta x$ or $\Delta y$ for information encoding. Fortunately, inverse design methods can overcome this issue. In Figure 2(d), we present an end-to-end inverse design algorithm based on a physical model $U_{\lambda,\hat{\sigma},z}(x,y) = f_{\lambda,z} \circ p_{\lambda,\hat{\sigma}}(\Delta\mathbf{r})$ that connects the position $\Delta\mathbf{r} = (\Delta x, \Delta y)$ of the meta-atoms to the scatted light field phase. Where $f_{\lambda,z}$ represents the free-space Fresnel diffraction which obviously deepens on both the wavelength $\lambda$ and the diffraction distance $z$. In the meantime, $p_{\lambda,\hat{\sigma}}$ dictates the metasurface modulated propagation-diffraction wave conversion process which depends on both the wavelength $\lambda$ and the incident guided wave direction $\hat{\sigma}$. More specifically, it utilizes $\Delta x$ and/or $\Delta y$ as the optimization parameters to directly find the meta-atom distribution for multi-wavelength applications. A differentiable forward model is essential for implementing this algorithm. According to Eq. (2), the relationship between the distribution of meta-atoms and the extracted field can be established

$$U_{\lambda,+\hat{x}}(i,j,0) = Ae^{j[\varphi_0 + \beta(\lambda)(\Lambda(i-1)+\Delta x)]}, \tag{4}$$

$$U_{\lambda,-\hat{x}}(i,j,0) = A'e^{j[\varphi_0' + \beta(\lambda)(\Lambda(M-i+1)-\Delta x)]}. \tag{5}$$

Note that here we have assumed identical scattered field amplitude $A$ and $A'$ for all
the meta-atoms by neglecting the decay of guide-wave intensity along the propagation direction since the total conversion efficiency is quite small (see Figure S2 in Supporting Information). In a more rigorous treatment, $A$ and $A'$ are position dependent and compensation by metaatom size design shall be considered.

The light field $U(i,j,z)$ on the parallel $xy$ plane at $z$ is calculated using the angular spectrum method (ASM) by two Fourier transformations $\mathcal{F}$ (see Supporting Information, Note

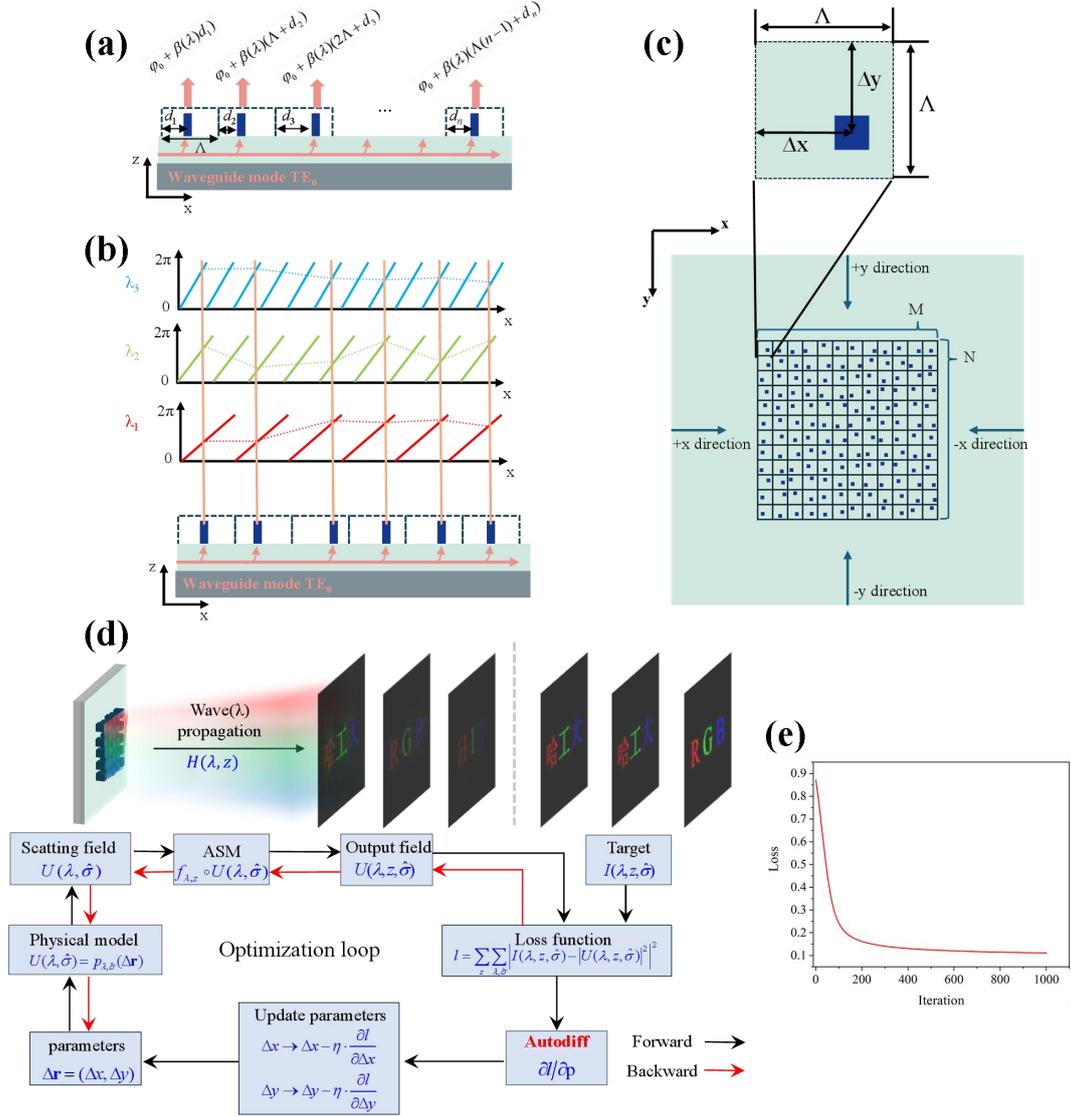

**Figure 2.** The detour phase and the scheme for target image reconstruction at multiple wavelengths. (a) Illustration depicting the phase of the extracted wave. (b) Illustration displaying the extracted wavefronts of the metasurface for multiple wavelengths. (c) Schematic of the metasurface comprising a pixel arrays. (d) End-to-end inverse design based on the physical model $U_{\lambda,\hat{\sigma},z}(x,y) = f_{\lambda,z} \circ p_{\lambda,\hat{\sigma}}(\Delta \mathbf{r})$ which connects the positions $(\Delta x, \Delta y)$ of metaatoms to the emitted light field phase (amplitude is approximately assumed identical). The light fields on the target plane are computed using the angular spectrum method. For the loop iteration, the loss function is derived from the target and reconstructed images, and the design parameters are updated via the gradient descent method to minimize the loss function. (e) Typical loss function $L$ converge curve.

S2 for details),

$$U(i,j,z) = \mathcal{F}^{-1}\{A(f_x,f_y,z)\} = \mathcal{F}^{-1}\{\{U_{\lambda,\hat{\sigma}}(i,j,0)\}H(f_x,f_y,z)\}. \qquad (6)$$

The loss function for the iteration loop is defined as the mean squared error between the target images $I(i,j,z_m)$, $m = 1,2,\ldots M$ and the corresponding images $|U_{\lambda,\hat{\sigma}}(i,j,z_m)|^2$ formed in the image planes $z_m$

$$L = \sum_{m=1}^{M} \sum_{\lambda,\hat{\sigma}} \left[ |U_{\lambda,\hat{\sigma}}(i,j,z_m)|^2 - I(i,j,z_m) \right]^2. \tag{7}$$

Note that Eq. (7) sums over all channels $k = \{\lambda, \hat{\sigma}\}$. Finally, the gradient descent algorithm is employed to adjust $\Delta \mathbf{r}$ ($\Delta x$ and/or $\Delta y$) and minimize the loss function based on Eqs. (4)-(7)

$$\frac{\partial L}{\partial \Delta \mathbf{r}} = \frac{\partial L}{\partial f} \frac{\partial f}{\partial p} \frac{\partial p}{\partial \Delta \mathbf{r}}, \tag{8}$$

$$\Delta \mathbf{r} \leftarrow \Delta \mathbf{r} - \eta \frac{\partial L}{\partial \Delta \mathbf{r}}. \tag{9}$$

More specifically, we build the forward model $U_{\lambda,\hat{\sigma},z}(x,y) = f_{\lambda,z} \circ p_{\lambda,\hat{\sigma}}(\Delta x, \Delta y)$ inside the Pytorch deep learning framework,[44] which enables automatic (Autodiff) differentiation and backpropagation (see Figure 2(d)). Figure 2(e) shows a typical iteration output curve for the case in Sect. 3.1 The iteration is stopped either for sufficient small loss function $L$ or iteration number.

It is noteworthy to remark that on-chip metasurfaces exploits the dispersion characteristics of the on-chip waveguide modes to generate a variety of phase profiles across different wavelength channels, utilizing a single positional arrangement of meta-atoms. Additionally, spatial dispersion takes place during the free-space diffraction process, allowing distinct wavelengths to create unique interference patterns within the same target plane. Our end-to-end inverse design approach utilizes these dispersion phenomena to achieve multiplexing across wavelength, diffraction distance, and incident direction channels. Remarkably, this methodology retains its effectiveness even in situations where the different wavelength channels are not completely decoupled--a considerable advantage over conventional forward design techniques, which often require such decoupling to ensure optimal performance.

## 3. Results and Discussions
### 3.1. Wavelength and Diffraction Distance Multiplexing

To verify the design strategy for multiwavelength operation, we optimized the position of the meta-atoms in the waveguide to achieve multi-plane RGB color holography, as shown in **Figure 3**(a). Here, we consider totally $900 \times 900$ pixels for the on-chip metasurface within a $360 \times 360$ μm$^2$ area, with $\Lambda$=400 nm. We incorporate three distinct imaging planes positioned at $z = 300$ μm, $z = 800$ μm, and $z = 1500$ μm, respectively. The metaatom and waveguide geometry is the same with that in Figure S1 in Supporting Information. We optimized the on-chip metasurface using the proposed method for simultaneous reconstruction of color holograms of Chinese characters "哈工大" (Chinese abbreviation for Harbin Institute of

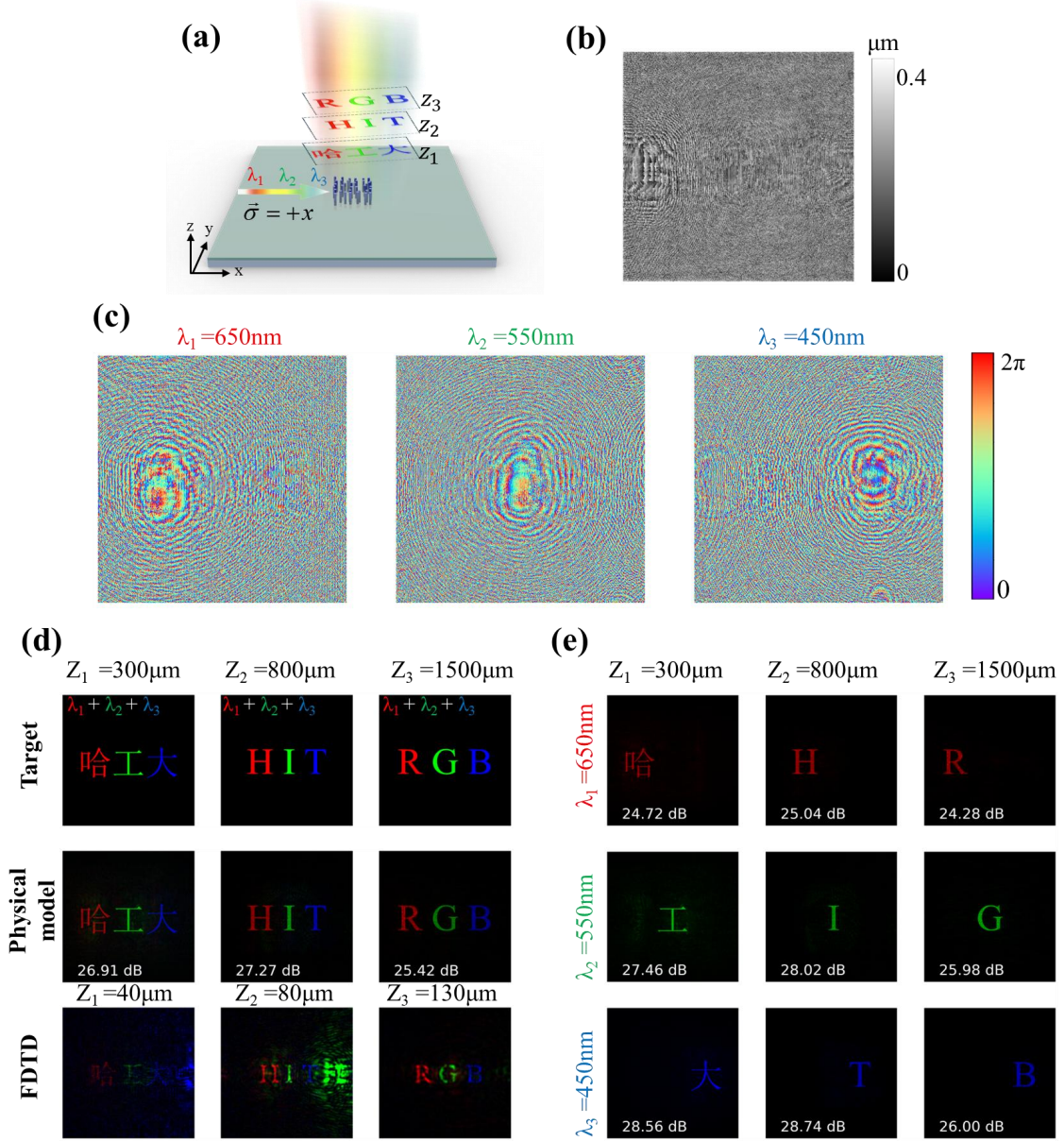

**Figure 3.** 9-channel theoretical physical model results and full-wave simulation results. (a) Schematic of the wavelength and diffraction distance multiplexing. (b) $\Delta x$ profile of the inverse-designed metasurface. (c) Phase profiles for $\lambda = 450$ nm, $\lambda = 550$ nm, and $\lambda = 650$ nm. (d) Target holograms and the physical model outputs for the plane at $z = 300$ μm, $z = 800$ μm, and $z = 1500$ μm, respectively. For comparison, FDTD simulation results are also shown, for the same on-chip metasurface with 150 × 150 pixels and reduced reconstruction distance (labelled respectively in the third row. (e) The individual holographic images output from the 9 holographic channels of the metasurface, respectively corresponding to the panels in second row in (d). These panels represent the physical model results, separated into color and distance channels.

Technology), English letters "HIT", and "RGB". Figure 3(b) shows the $\Delta x$ map of the finally optimized metasurface by the scheme shown in Figure 2, and Figure 3(c) shows the phase map at $\lambda = 450$ nm, $\lambda = 550$ nm, and $\lambda = 650$ nm, respectively. As seen in Figure 3(d) and 3(e), the reconstructed image at each $z$ plane closely matches the target hologram. Figure 3(e) illustrates the observation of each individual channel as shown in the middle row of Figure 3(d), with the color and distance channels distinctly separated for enhanced clarity. Simultaneously, we computed the PSNR of the holographic images, where PSNR is defined as PSNR=$10 \log(1/MSE)$ and $MSE$ represents the mean square error between the reconstructed holographic image and the target image. The PSNR of each holographic image is indicated in the lower left corner. In addition, we verified the effectiveness of the optimized metasurface by full-wave simulation with the finite-difference time-domain (FDTD) method (see Supporting Information, Note S3 for details). The simulation results are also shown in Figure 3(d). Note that, we reduce the metasurface to 150 × 150 pixels for the FDTD full-wave simulation, considering the computational memory and speed limitations. Note further that the diffraction distance is also reduced for the FDTD simulation. In this regard, for the size-reduced metasurface, the designed $\Delta x$ map is independently optimized (see Supporting Information, Note S3 for details), concerning the original large-size metasurface of 900 × 900 pixels. Nevertheless, the 3D full-wave numerical results agree quite well with the target one and the physical model outputs.

### 3.2. Wavelength, Diffraction Distance, and Guided Wave Direction Multiplexing

In this subsection, we use the proposed end-to-end inverse design method to design a metasurface with incident direction multiplexing of the guided wave excitation. Different holographic images are projected as the guided wave propagates along different directions, as shown in **Figure 4**(a). We keep the target images in the $+\hat{x}$ direction consistent with those in Figure 3(d) and set the target images for the $-\hat{x}$ direction as objects of a RGB palette, a magic cube, and a flower. Notice that according to Eqs. (4) and (5), the $+\hat{x}$ and $-\hat{x}$ direction channels are not totally decoupled. Figure 4(b) shows the multi-plane RGB color hologram reconstruction results for the $+\hat{x}$ and $-\hat{x}$ guided wave excitation, respectively. In the $+\hat{x}$ excitation direction, the reconstructed holograms exhibit a higher background noise and a lower signal-to-noise ratio compared to those in Figure 4(b). This can be ascribed to the fact that we double the holographic channels (9 × 2=18) while still relying solely on the same design space $\Delta x$ of meta-atoms as the encoding degrees of freedom. We also verified this guided wave excitation direction multiplexing using FDTD simulation (see Supporting Information Note S4 for details). It is natural to consider more direction multiplexing and guided waves can be excited in four directions: $\pm\hat{x}$ and $\pm\hat{y}$. Moreover, by multiplexing these four directions, we can design both $\Delta x$ and $\Delta y$ maps and can finally obtain $9\times 4 = 36$ holographic images, forming a keyboard-like pattern, as shown in Figure 4(c).

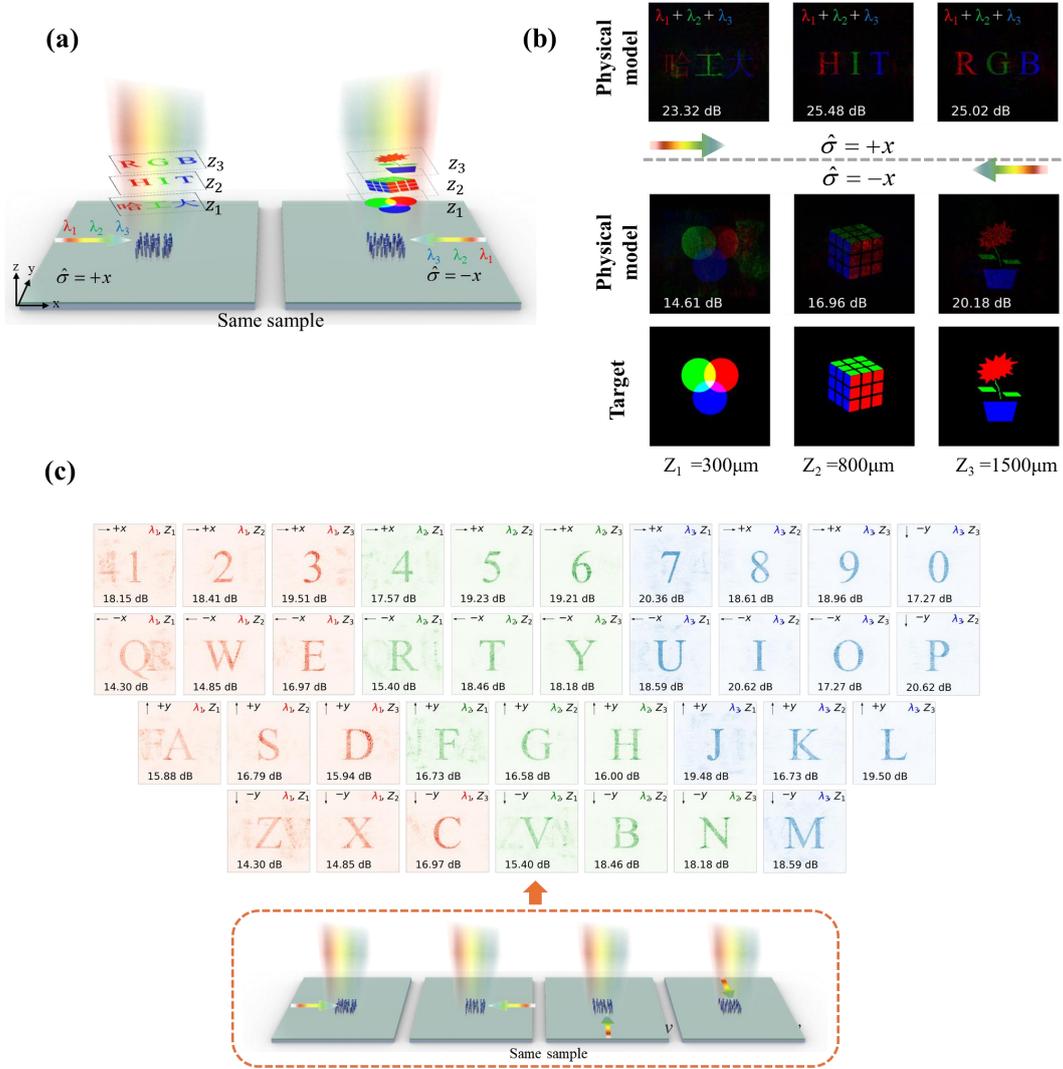

**Figure 4.** Propagation direction multiplexed hologram with 18 channels and 36 channels, respectively. (a) Schematic of the guided wave direction multiplexing. The guided waves injected from $\pm\hat{x}$ directions can generate different RGB holographic images at multiple planes. (b) The physical model outputs of 18 holographic images agree with the target images for the bi-directional case. (c) A keyboard pattern consisting of 36 holograms formed by multiplexing the four directions of $\pm\hat{x}$ and $\pm\hat{y}$.

### 3.3. Diffraction Distance Multiplexing Co-Design

To show the versatility of our method, we use the proposed algorithm for the co-design of two orthogonal excitation directions $\pm\hat{x}$ and $\pm\hat{y}$. In other words, it is possible to enhance the final projected images by setting the same target for different direction channels. More specifically, the metaatom distributions along the $x$ and $y$-axes can scatter lights off the chip and map them to the same final images. For this case, we designed 10 target planes [see **Figure 5**(a)], each corresponding to a digital tube number '0'-'9', as shown in Figure 5(b). For comparison, we also design a metasurface for this task using only $\Delta x$ as the optimization parameter and show the results in Figure 5(c). Figure 5(d) shows the co-design results using both $\Delta x$ and $\Delta y$ as the optimization parameters. It is seen that the holographic images quality has improved significantly with respect to those in Figure 5(c). The comparison of PSNR values is shown in Figure 5(e). It is clearly seen that the co-design strategy increased the PSNR by about 1.5 dB. We stress that for reduced target plane number, the co-design hologram visibility enhancement effect is more dramatic. An additional example for 5 target planes is provided in Figure S8 in see Supporting Information. Co-design has the potential to enhance the quality of holographic images. However, its impact is limited in some channels due to pronounced crosstalk between different target planes, as illustrated by the holographic images labeled '6, 7, 8' in Figure 5. To mitigate this issue, one could integrate a penalty term into the loss function and/or utilize non-uniformly spaced target planes to increase the separation between planes that exhibit substantial crosstalk.

It is crucial to emphasize that, given the channels being not decoupled, there exists a trade-off between the number of imaging planes and the signal-to-noise ratio of the holographic image. Our proposed end-to-end design scheme identifies the optimal solution subjected to this trade-off.

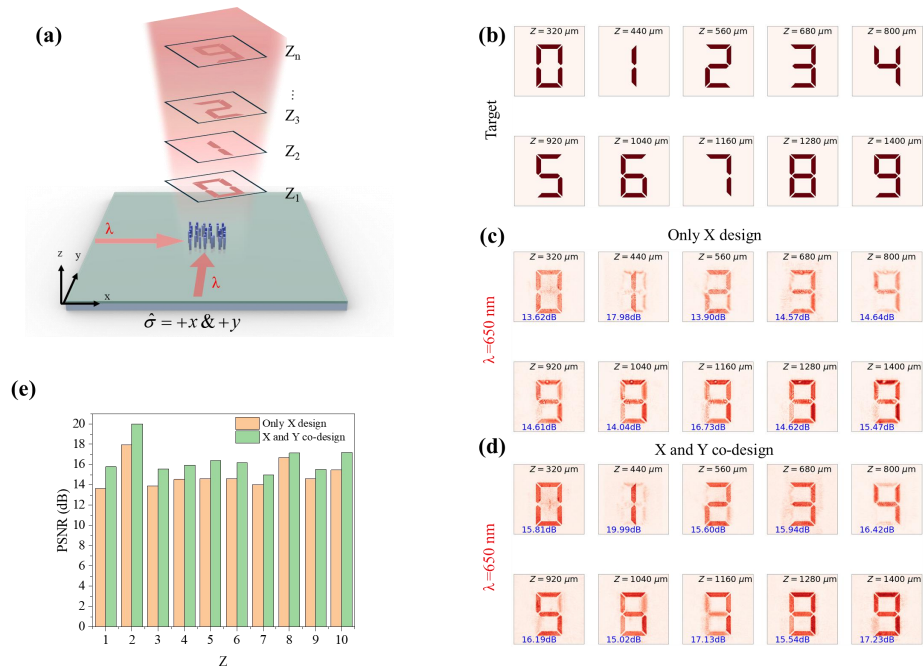

**Figure 5.** Holograms by co-design. (a) Schematic of the co-design method with guided wave coming from two geometrically orthogonal directions. (b) The target images of digital number. (c) The results solely using $\Delta x$ as the design parameter. (d) The results using both $\Delta x$ and $\Delta y$ as the design space parameters. (e) Comparison of the peak signal-to-noise ratio (PSNR) of the holographic images resulted by the two design methods.

## 4. Conclusions

In this study, we introduce an efficient inverse design framework specifically tailored for on-chip metasurfaces, aimed at multi-dimensional holographic applications. This approach diverges from conventional forward methods and electromagnetic solver-embedded inverse design strategies by rapidly producing metasurface designs. We substantiate the accuracy of our physical model through rigorous 3D FDTD simulations, thereby validating the effectiveness of our framework.

In contrast to free-space metasurfaces, on-chip metasurfaces achieve phase modulation by altering the positions of meta-atoms within a single pixel. This approach, however, may increase the metasurface's vulnerability to fabrication errors. To address this, we conducted a comprehensive assessment of the impact of fabrication errors on hologram quality through both full-wave simulations and theoretical model predictions (for further details, refer to the Supporting Information, Note 6). The findings confirm the metasurface's robustness against fabrication errors up to ±20 nm (5% of a pixel size of 400 nm), which is within the acceptable tolerance range of electron beam lithography and standard focused ion beam techniques.[45,46]

Our proposed end-to-end inverse design method has successfully implemented holography with wavelength-channel multiplexing. However, the fidelity of real-color reproduction falls short, as indicated by the RGB palette in Figure 4(b), which shows elevated noise levels—a common challenge faced by pure phase holograms. To enhance full-color holography, incorporating complex-amplitude modulation is a strategic solution. Within our design framework, the realization of real-color holography is feasible by developing a physical model for the meta-atom that supports complex-amplitude modulation.

By harnessing the distinctive properties of guided-wave-driven metasurfaces, we aim to contribute to the advancement of on-chip optical devices and develop new opportunities in integrated photonics. Our inverse design approach, grounded in physical model, offers an advantage over conventional forward design methods by enabling the handling of complex tasks and the creation of hypermultiplexed channels, considering factors like wavelength, diffraction distance, and the incoming direction of guided waves. This framework allows accommodation of more multiplexing dimensions in terms of guided waveguide mode type (e.g., TE or TM mode[47]), polarization,[48] and on-chip source excitation (e.g., by quantum dots[49,50]).

The integration of metasurfaces with waveguides is expected to pave the way for the development of compact, high-performing optical components. This integration will facilitate innovative applications in off-chip beam profile control and promote advancements in optical information processing and computing. However, it is important to acknowledge the inherent limitations in the overall energy efficiency. The multi-objective optimization method we propose is applicable to a broad range of scenarios, allowing for increased flexibility and the ability to manipulate additional dimensions of the light field, such as trajectory, and angular momentum. Additionally, the metasurface structures designed through this method are ready for fabrication using techniques like electron-beam lithography, laser additive manufacturing, direct laser writing or nano-imprinting.[51-53]

**Supporting Information**

Supporting Information is available from the Wiley Online Library or from the author.

**Acknowledgements**

This work was supported by the National Natural Science Foundation of China (No. 62375064), Shenzhen Science and Technology Program (No. JCYJ20210324132416040, KJZD20230923114803007), Guangdong Provincial Nature Science Foundation (No. 2022A1515011488), the National Key Research and Development Program of China (No. 2022YFB3603204), and Guangdong Basic and Applied Basic Research Foundation (No. 2023A1515110572).

**Conflict of Interest**

The authors declare no conflict of interest.

**Data Availability Statement**

The data that support the findings of this study are available from the corresponding author upon reasonable request.

References

[1] A. Ma´rquez, C. Li, A. Bel´endez, S. A. Maier, H. Ren, *Nanophotonics* **2023**, *12*, 4415.

[2] B. Ko, T. Badloe, Y. Yang, J. Park, J. Kim, H. Jeong, C. Jung, J. Rho, *Nat. Commun.* **2022**, *13*, 6256.

[3] S. So, J. Mun, J. Park, J. Rho, *Adv. Mater.* **2023**, *35*, 2206399.

[4] Y. Hu, X. Luo, Y. Chen, Q. Liu, X. Li, Y. Wang, N. Liu, H. Duan, *Light: Sci. Appl.* **2019**, *8*, 86.

[5] I. Kim, W.-S. Kim, K. Kim, M. A. Ansari, M. Q. Mehmood, T. Badloe, Y. Kim, J. Gwak, H. Lee, Y.-K. Kim, J. Rho, *Sci. Adv.* **2021**, *7*, eabe9943.

[6] S. So, J. Kim, T. Badloe, C. Lee, Y. Yang, H. Kang, J. Rho, *Adv. Mater.* **2023**, *35*, 2208520.

[7] X. Liu, D. Zhang, L. Wang, T. Ma, Z. Liu, J.-J. Xiao, *Photonics* **2023**, *10*, 503.

[8] S. Chen, W. Liu, Z. Li, H. Cheng, J. Tian, *Adv. Mater.* **2020**, *32*, 1805912.

[9] G. Kim, S. Kim, H. Kim, J. Lee, T. Badloe, J. Rho, *Nanoscale* **2022**, *14*, 4380.

[10] D. Wen, J. J. Cadusch, J. Meng, K. B. Crozier, *Adv. Photonics* **2021**, *3*, 024001.

[11] Y. Chen, X. Zheng, X. Zhang, W. Pan, Z. Wang, S. Li, S. Dong, F. Liu, Q. He, L. Zhou, S. Sun, *Nano Lett.* **2023**, *23*, 3326.

[12] F. Ding, S. I. Bozhevolnyi, *IEEE J. Sel. Top. Quantum Electron.* **2019**, *25*, 1.

[13] L. Huang, X. Chen, B. Bai, Q. Tan, G. Jin, T. Zentgraf, S. Zhang, *Light: Sci. Appl.* **2013**, *2*, e70.

[14] Y. Liu, Y. Shi, Z. Wang, Z. Li, *ACS Photonics* **2023**, *10*, 1268.

[15] K. Liao, T. Gan, X. Hu, Q. Gong, *Nanophotonics* **2020**, *9*, 3315.

[16] Z. Wang, L. Chang, F. Wang, T. Li, T. Gu, *Nat. Commun.* **2022**, *13*, 2131.

[17] T. Fu, Y. Zang, Y. Huang, Z. Du, H. Huang, C. Hu, M. Chen, S. Yang, H. Chen, *Nat. Commun.* **2023**, *14*, 70.

[18] Y. Ha, Y. Guo, M. Pu, X. Li, X. Ma, Z. Zhang, X. Luo, *Adv. Theory Simul.* **2021**, *4*, 2000239.

[19] Y. Ding, X. Chen, Y. Duan, H. Huang, L. Zhang, S. Chang, X. Guo, X. Ni, *ACS photonics* **2022**, *9*, 398.

[20] Z. Liu, D. Wang, H. Gao, M. Li, H. Zhou, C. Zhang, *Adv. Photonics* **2023**, *5*, 034001.

[21] S. Yu, L. Ranno, Q. Du, S. Serna, C. McDonough, N. Fahrenkopf, T. Gu, J. Hu, *Laser Photonics Rev.* **2023**, *17*, 2200025.

[22] G. Xu, A. Overvig, Y. Kasahara, E. Martini, S. Maci, A. Alu`, *Nat. Commun.* **2023**, *14*, 4380.

[23] P.-Y. Hsieh, S.-L. Fang, Y.-S. Lin, W.-H. Huang, J.-M. Shieh, P. Yu, Y.-C. Chang, *Nanophotonics* **2022**, *11*, 4687.

[24] B. Fang, Z. Wang, S. Gao, S. Zhu, T. Li, *Nanophotonics* **2022**, *11*, 1923.

[25] Y. Shi, C. Wan, C. Dai, S. Wan, Y. Liu, C. Zhang, Z. Li, *Optica* **2022**, *9*, 670.

[26] Y. Zhang, Z. Li, W. Liu, H. Cheng, D.-Y. Choi, J. Tian, S. Chen, *Laser Photonics Rev.* **2023**, *17*, 2300109.

[27] J. Ji, Z. Wang, J. Sun, C. Chen, X. Li, B. Fang, S.-N. Zhu, T. Li, *Nano Lett.* **2023**, *23*, 2750.

[28] P.-Y. Hsieh, S.-L. Fang, Y.-S. Lin, W.-H. Huang, J.-M. Shieh, P. Yu, Y.-C. Chang, *Opt. Express* **2023**, *31*, 12487.

[29] S. Chen, J. Huang, S. Yin, M. M. Milosevic, H. Pi, J. Yan, H. M. Chong, X. Fang, *Opt. Express* **2023**, *31*, 15876.

[30] Y. Yin, Q. Jiang, H. Wang, J. Liu, Y. Xie, Q. Wang, Y. Wang, L. Huang, *Adv. Mater.* **2024**, 2312303.

[31] J. Kim, J.-H. Im, S. So, Y. Choi, H. Kang, B. Lim, M. Lee, Y.-K. Kim, J. Rho, *Adv. Mater.* 2024, 2311785.

[32] B. Xiong, Y. Liu, Y. Xu, L. Deng, C.-W. Chen, J.-N. Wang, R. Peng, Y. Lai, Y. Liu, M. Wang, *Science* **2023**, *379*, 294.

[33] Z. Xie, Z. Liang, H. Wu, Q. Zeng, Z. Guan, A. Long, P. Zhong, J. Liu, H. Ye, D. Fan, S. Chen, *Nanophotonics* **2024**, *13*, 529.

[34] Z. Huang, Y. He, P. Wang, W. Xiong, H. Wu, J. Liu, H. Ye, Y. Li, D. Fan, S. Chen, *Opt. Express* **2022**, *30*, 5569.

[35] H. Ren, G. Briere, X. Fang, P. Ni, R. Sawant, S. Heron, S. Chenot, S. Vezian, B. Damilano, V. Brandli, S. A. Maier, P. Genevet, *Nat. Commun.* **2019**, *10*, 2986.

[36] Y. Yang, J. Seong, M. Choi, J. Park, G. Kim, H. Kim, J. Jeong, C. Jung, J. Kim, G. Jeon, K. Lee, D. H. Yoon, J. Rho, *Light: Sci. Appl.* **2023**, *12*, 152.

[37] J. Guo, Y. Zhang, H. Ye, L. Wang, P. Chen, D. Mao, C. Xie, Z. Chen, X. Wu, M. Xiao, Y. Zhang, *ACS Photonics* **2023**, *10*, 757.

[38] Z. Li, Y. Liu, C. Zhang, Y. Qiao, R. Deng, Y. Shi, Z. Li, *Adv. Funct. Mater.* **2023**, 2312705.

[39] R. Yang, S. Wan, Y. Shi, Z. Wang, J. Tang, Z. Li, *Laser Photonics Rev.* **2022**, *16*, 2200127.

[40] Y. Shi, C. Wan, C. Dai, Z. Wang, S. Wan, G. Zheng, S. Zhang, Z. Li, *Laser Photonics Rev.* **2022**, *16*, 2100638.

[41] B. Fang, F. Shu, Z. Wang, J. Ji, Z. Jin, Z. Hong, C. Shen, Q. Cheng, T. Li, *Opt. Lett.* **2023**, *48*, 3119.

[42] D. Zhang, Z. Liu, X. Yang, J. J. Xiao, *ACS Photonics* **2022**, *9*, 3899.

[43] X. Guo, Y. Ding, X. Chen, Y. Duan, X. Ni, *Sci. Adv.* **2020**, *6*, eabb4142.

[44] PyTorch, https://pytorch.org/.

[45] Y. Chen, *Microelectron. Eng.* 2015, *135*, 57.

[46] O. Buchnev, J. A. Grant-Jacob, R. W. Eason, N. I. Zheludev, B. Mills, K. F. MacDonald, *Nano Lett.* 2022, *22*, 2734

[47] K. Zhao, Y. Ha, Y. Guo, M. Pu, Y. Fan, X. Li, F. Zou, L. Wang, S. She, X. Luo, *Adv. Opt. Mater.* 2303009.

[48] B. Fang, Z. Wang, Y. Li, J. Ji, K. Xi, Q. Cheng, F. Shu, Z. Jin, Z. Hong, C. Zhan, C. Shen, T. Li, *Photonics Res.* **2023**, *11*, 2194.

[49] Y. Kan, X. Liu, S. Kumar, S. I. Bozhevolnyi, *ACS Photonics* **2024**. *11*, 1584.

[50] M. Jeong, B. Ko, C. Jung, J. Kim, J. Jang, J. Mun, J. Lee, S. Yun, S. Kim, J. Rho, *Nano Lett.* 2024, *24*, 5783.

[51] H.-H. Hsiao, C. H. Chu, D. P. Tsai, *Small Methods* **2017**, *1*, 1600064.

[52] G. Yoon, T. Tanaka, T. Zentgraf, J. Rho, *J. Phys. D-Appl. Phys.* **2021**, *54*, 383002.

[53] H. Ren, X. Fang, J. Jang, J. B urger, J. Rho, S. A. Maier, *Nat. Nanotechnol*. 2020, *15*, 948.

# Supporting Information

# Hypermultiplexed off-chip hologram by on-chip integrated metasurface

Xianjin Liu,[1,2] Zhanying Ma,[1,2] Dasen Zhang,[1,2] Qiwen Bao,[1,2] Zhenzhen Liu,[1,2,3] and Jun-Jun Xiao[1,2,4,*]

[1]*College of Electronic and Information Engineering, Harbin Institute of Technology (Shenzhen), Shenzhen 518055, China*

[2]*Shenzhen Engineering Laboratory of Aerospace Detection and Imaging, Harbin Institute of Technology (Shenzhen), Shenzhen 518055, China*

[3]*Research Center for Advanced Optics and Photoelectronics, Department of Physics, College of Science, Shantou University, Shantou 515063, China*

[4]*College of Integrated Circuits, Harbin Institute of Technology (Shenzhen), Shenzhen 518055, China*

# Contents

## Supplementary Note 1: 3D FDTD simulation and beam focusing metasurface designed for a single wavelength

Figure S1(a) illustrates the meta-atom structure made of amorphous silicon (α-Si) square pillar positioned on a 220 nm-thick $Si_3N_4$ waveguide, with the square pillar side length $L = 100$ nm and height $H = 380$ nm. As a proof-of-concept demonstration of our model and justification by the numerical calculations, we position the meta-atoms on a strip waveguide to create a 1D on-chip metalens for focusing the scattered wave into a point in the free space, as shown in Figure S1(b). The strip waveguide is of width $w = 500$ nm and the on-chip metalens is exclusively illuminated by the fundamental transverse electric mode (TE00) that has dominate $E_y$ component. Figure S1(c) shows the dependence of the TE00 mode propagation constant $\beta$ on the wavelength $\lambda$. In the 1D metalens scenario, the red line in Figure S1(c) denotes $\beta(\lambda)$ in the strip waveguide. Conversely, in the off-chip holography scenario, the gray line in Figure S1(c) represents $\beta(\lambda)$ in a slab waveguide. Typical mode profiles for strip waveguide and slab waveguide are shown as the insets in Figure S1(c). Note that here we have designed the metalens operating at wavelength $\lambda$=500 nm. For this particular wavelength, the phase variation of the TE00 mode is represented by the blue curve in Figure S1(d), periodically wrapped per $2\pi$. To obtain the required phase [see the yellow curve in Figure S1(b)] of the extracted wave $\varphi(x) = -k_0\sqrt{x^2 + f^2} - f$, which focuses towards the point (0, $f$=6 μm), the position coordinates of the meta-atoms are determined by the intersection of these two curves. The 3D FDTD simulation results are depicted in Figure S1(e). It is seen that for wavelength $\lambda$=650 nm, the off-chip scattered wave is well-focused at the preset spot. However, for other wavelengths, the focus spot shifts along the $x$-axis due to changes in the propagation constant. This implies that at different wavelengths, the TE00 mode interacting with the metasurface can lead to different holographic images on the target plane, by deliberate design.

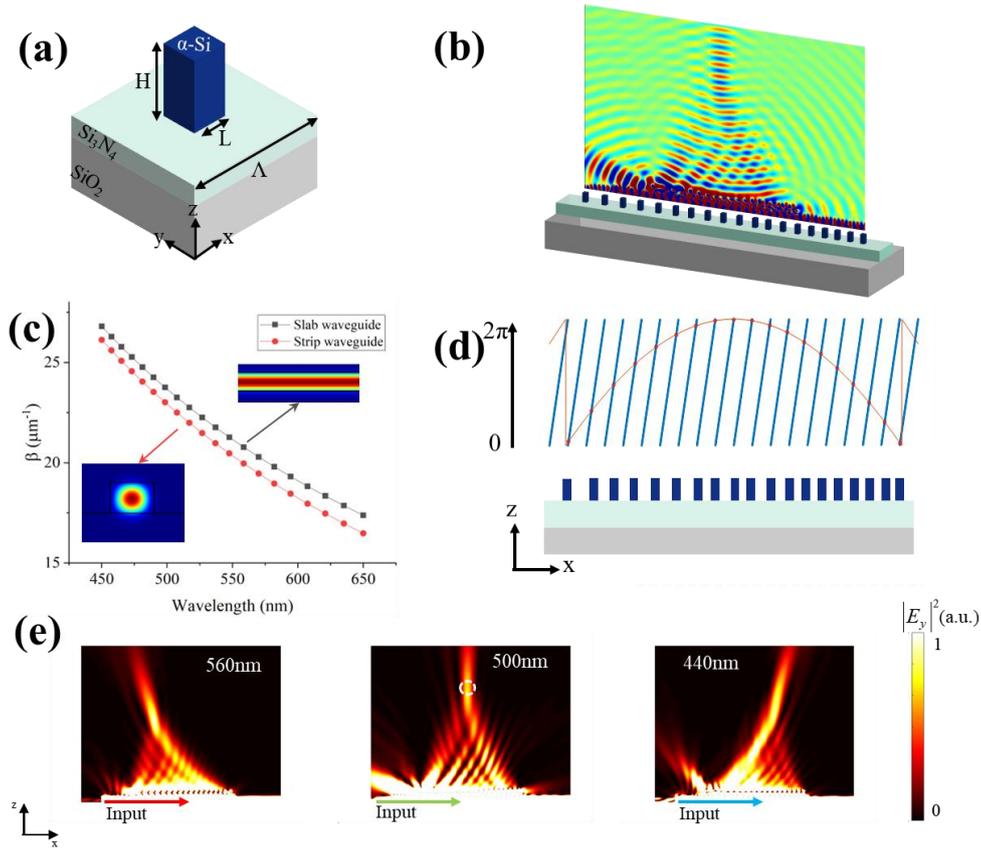

**Figure S1.** (a) Geometric structure of the meta-atom, consisting of a α-Si square pillar of height $H = 380$ nm and side length $L = 100$ nm, arranged on a Si3N4 slab waveguide. (b) 1D on-chip metalens with 20 metaatoms that focuses the off-chip scattered waves into a point in the free space (marked by the white circle) 6μm above the metasurface. (c) The propagation constant dispersion $\beta(\lambda)$ of the fundamental TE0 mode in a strip waveguide ($w = 500$ nm) and a slab waveguide ($w = \infty$). (d) Diagram illustrating the strategy of finding the positions of meta-atom along the $x$-axis to form a 1D metalens. The blue curve depicts the transmission phase of the TE0 mode, while the orange curve represents the required/target phase for focusing the off-chip scattered wave. The intersections of two curves are used to determine the position of the meta-atoms. (e) Full-wave simulation of a 1D metalens for off-chip focusing. The target focusing point is marked by the white circles.

Figure S2 shows that the maximum phase accumulation inside each metaatom satisfies $\Lambda\beta(\lambda) > 2\pi$ for the slab waveguide. Here the metasurface pixel size is $\Lambda = 400$ nm. The insets show the FDTD simulation results for the propagation of guided wave in presence of a metasurface with $40 \times 40$ metaatoms. It is seen that the guide-wave decay is quite small, validating the assumption of constant scattering amplitude as used in Eqs. (4) and (5) in main text.

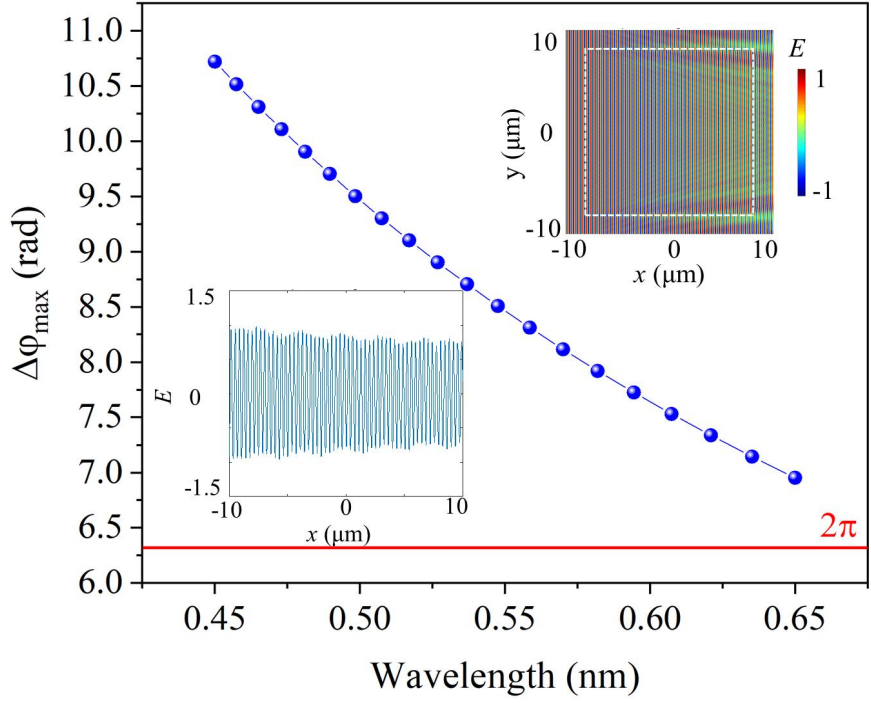

**Figure S2.** Guided wave phase accumulation $\Delta\varphi_{max}$ across a metasurface pixel for all interested operating wavelength. The insets show the off-chip scattering (for $\lambda$=650 nm) leads to negligible effect on the guided wave amplitude across the metasurface region.

## Supplementary Note 2: Raleigh-Sommerfeld diffraction model

To meet the computational speed and accuracy for the inverse design, we employ the angular spectrum method (ASM), to calculate the light field in the spatial-frequency domain by treating the light field as a combination of plane waves traveling in various directions. This propagation is linear and can be expressed as

$$A(f_x, f_y, z) = A(f_x, f_y, 0) H(f_x, f_y, z). \tag{10}$$

Here, the angular spectrum $A(f_x, f_y, z)$ of the light field $U(x, y, z)$, and the transfer function $H(f_x, f_y, z)$, are respectively given by

$$A(f_x, f_y, z) = \iint U(x, y, z) e^{-j2\pi(xf_x + yf_y)} dx dy, \tag{11}$$

and

$$H(f_x, f_y, z) = \begin{cases} \exp[jkz\sqrt{1-(\lambda f_x)^2 - (\lambda f_y)^2}], & f_x^2 + f_y^2 < \dfrac{1}{\lambda} \\ 0, & \text{other} \end{cases}, \tag{12}$$

Where $\mathcal{F}$ represents the Fourier transform and $f_x$ and $f_y$ denote the spatial frequencies along the $x$ and $y$ directions, respectively. Consequently, the light field $U(x, y, z)$ is derived through the inverse Fourier transform:

$$U(x, y, z) = \mathcal{F}^{-1}(A(f_x, f_y, z)) = \mathcal{F}^{-1}(A(f_x, f_y, 0) H(f_x, f_y, z)). \tag{13}$$

For wavelength-multiplexing on-chip metasurface applications, we assign blue (B), green (G), and red (R) colors to wavelengths of $\lambda$ = 450 nm, 550 nm, and 650 nm, respectively. To satisfy the Nyquist sampling criterion for the Fourier transform, we incorporate sufficient zero padding to prevent aliasing errors.

# Supplementary Note 3: 3D FDTD simulation for the case of wavelength and diffraction distance multiplexing

We utilize FDTD simulation package to validate the effectiveness of the on-chip metasurface designed for wavelength and distance multiplexing. With limited available computing resources, we have used a metasurface of $150 \times 150$ pixels and set the three target planes at $z = 40$ μm, $z = 80$ μm, and $z = 130$ μm. Firstly, the metasurface is optimized using the proposed method. Figure S3(a) shows the $\Delta x$ map of the optimized metasurface, and Figure S3(b) shows the physical structure. Figure S3(c) exhibits the ideal phase maps by the theoretical mode for $\lambda = 450$ nm, $\lambda = 550$ nm, and $\lambda = 650$ nm, respectively. Next, we conduct the FDTD simulation using the optimized $\Delta x$ map. The field monitor is placed 2 μm above the metasurface to capture the scattering fields. Then, the ASM is utilized to compute the holographic images in the target planes using the FDTD-extracted light fields. Compared to the ideal phase maps (Figure S3(c)), the phase maps of the extracted light field obtained in the monitor plane (see Figure S3(d)) matches the best for $\lambda = 650$ nm. and for $\lambda = 550$ nm and $\lambda = 450$ nm, some high-frequency noise is present. This is attributed to the low-precision grid, leading to less accurate simulation results at shorter wavelengths. Figure S3(e) and S3(f) show the holographic images from the FDTD simulation and the physical model, respectively. It is clearly seen that the holograms from the FDTD simulation have lower quality compared to those from the physical model. Nonetheless, they remain highly distinguishable, with all Chinese characters and English letters visible. These results demonstrate the effectiveness of the designed on-chip metasurfaces with wavelength and distance multiplexing.

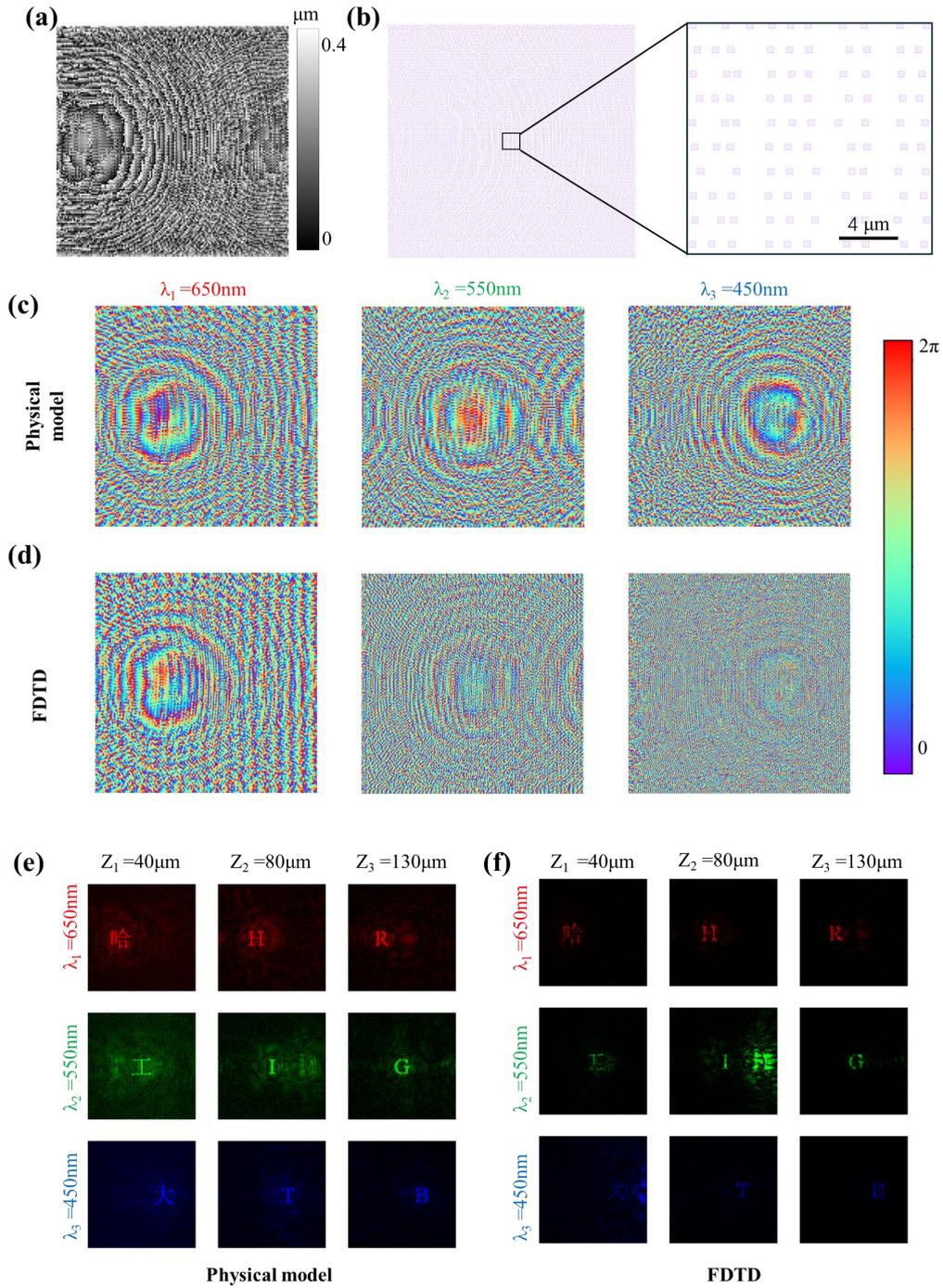

**Figure S3.** FDTD simulation results. (a) The $\Delta x$ map of the optimized metasurface. (b) The structure (GDSII) of the optimized metasurface, consisting of $150 \times 150$ pixels, in an area of $60 \times 60$ μm². Phase maps from (c) the physical model and (d) the FDTD simulation for $\lambda = 650$ nm, $\lambda = 550$ nm, and $\lambda = 450$ nm, respectively. The 9 individual holographic images from (e) the physical model and (f) from the FDTD simulation.

# Supplementary Note 4: Optimized results for the case of wavelength, diffraction distance, and guided wave direction multiplexing

Figure S4 shows more results on the case of multiplexing in terms of wavelength, diffraction distance, and guided wave direction. In Figure S4(a), the $\Delta x$ map of the optimized metasurface is shown. Figure S4(b) and S4(c) present the phase maps for the $+\hat{x}$ and $-\hat{x}$ guided wave excitation, respectively, at wavelengths of $\lambda = 650$ nm, $\lambda = 550$ nm, and $\lambda = 450$ nm. For one specific wavelength, the phase maps for the $+\hat{x}$ and $-\hat{x}$ direction excitation exhibit similar fringes, which arises from the two phase map relationship $\varphi_{-x}(i,j) = \varphi_c(\lambda) - \varphi_{+x}(i,j)$ where $\varphi_c = 2\varphi_0 + \beta(\lambda)M\Lambda$, $M$ is the $x$-direction resolution size. Figure S4(d) and S4(e) show the 9 holographic images for the $+\hat{x}$ and $-\hat{x}$ guided wave excitation, respectively. The results suggest that for more holographic channels with a fixed geometric coding degree of freedom, the crosstalk could increase and holographic image quality degrades, as expected.

    We also verify the guided wave direction multiplexing of the on-chip metasurfaces using 3D FDTD simulations. To maintain a reasonable hologram computation time, we reduce the size of the metasurface to $40 \times 40$ pixels within a $60 \times 60$ μm2 area and operates at single wavelength $\lambda = 650$ nm, for one specific diffraction distance $z = 30$ μm. As shown in Figure S5(a), the target holographic images for the two guided wave directions, $+\hat{x}$ and $-\hat{x}$, are set as the letters "X" and "J", respectively. The $\Delta x$ map of the optimized metasurface is shown in Figure S5(b). Figure S5(c) shows the phase maps of the scattered light field by the metasurface under $+\hat{x}$ and $-\hat{x}$ guided wave excitations. The phase maps obtained from the FDTD simulation closely agree with the ideal phase maps from the physical model. Figure S5(d) shows the reconstructed holographic images: with $+\hat{x}$ guided wave excitation the metasurface projects the symbol "X", while with $-\hat{x}$ excitation it generates the letter "J".

    Figure S6 shows the optimized structure details for the keyboard patterns shown in Figure 4(c) in the main text.

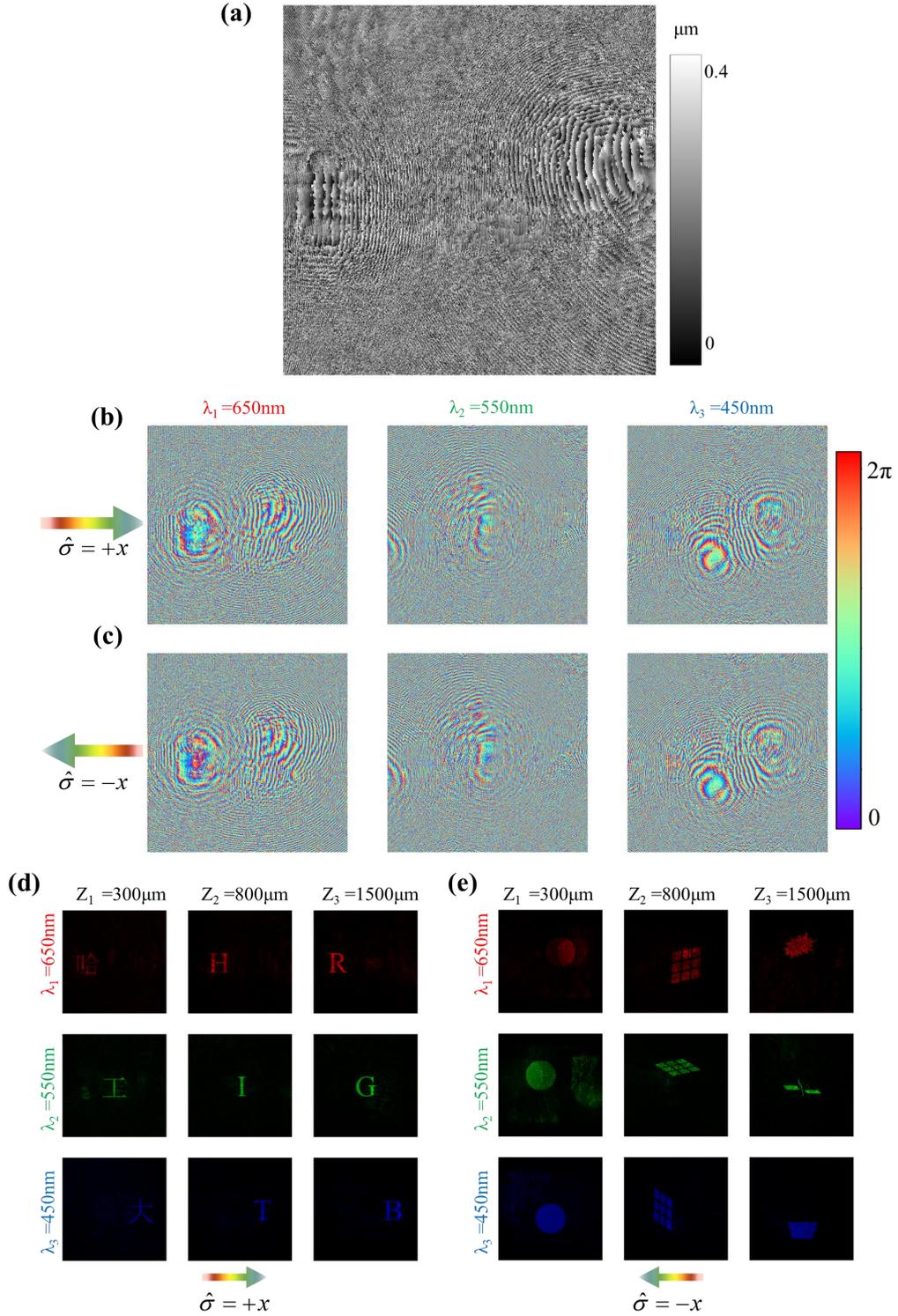

**Figure S4.** More results regarding wavelength, diffraction distance, and guided wave direction multiplexing. (a) The $\Delta x$ map of the optimized metasurface. (b) and (c) depict phase maps under $+\hat{x}$ and $-\hat{x}$ guided wave excitations, respectively, for wavelength $\lambda$ = 450 nm, 550 nm, and 650 nm. (d) and (e) showcase 9 individual holographic images each, corresponding to $+\hat{x}$ and $-\hat{x}$ guided wave excitation, respectively.

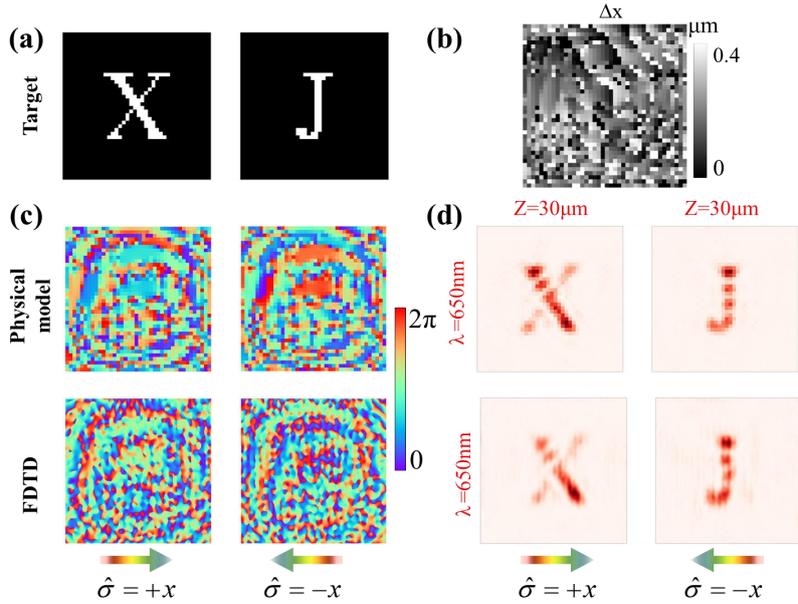

**Figure S5.** FDTD simulation of the guided wave direction multiplexing. (a) The target holograms generated by guided wave coming from two opposite directions $+\hat{x}$ and $-\hat{x}$. (b) The $\Delta x$ map of the optimized metasurface. (c) Phase maps from physical model (left column) and from FDTD simulation (right column), under the excitation of guided wave from opposite directions, $+\hat{x}$ and $-\hat{x}$. (d) Generated holographic images at the receiving plane, closely resembling the target ones.

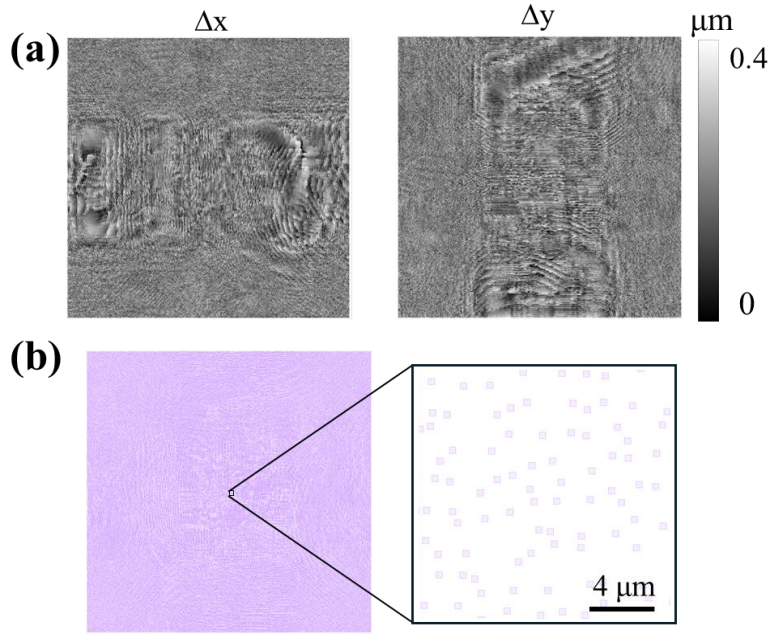

**Figure S6.** On-chip metasurface designed for multiplexing of four guide-wave directions. (a) The $\Delta x$ map (left) and $\Delta y$ map (right) of the optimized metasurface. (b) Structure layout of the optimized metasurface and a zoom-in part.

# Supplementary Note 5: Optimized results for the co-designed cases of guided wave excitation from orthogonal directions

Figure S7 shows the optimized structure and bi-directional phase maps for the case of 10 digital numbers shown in Figure 5 in the main text. Figure S8 shows the holograms by the co-design method for 5 target planes.

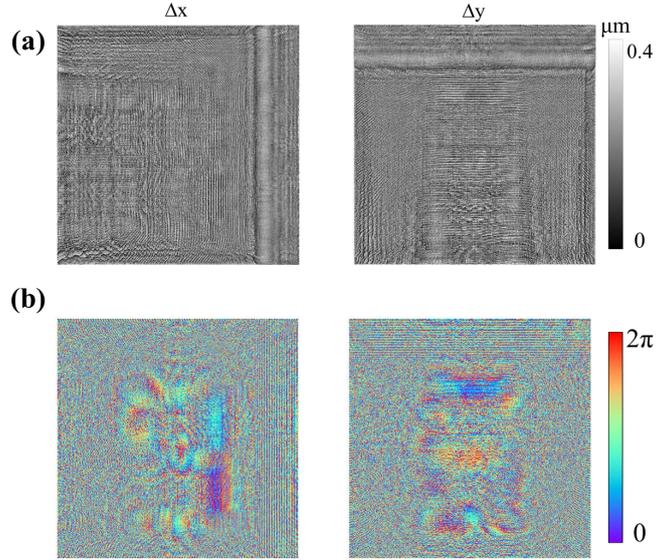

**Figure S7.** Co-design of guided-wave excitation simultaneously from two orthogonal directions. (a) The $\Delta x$ map (left) and $\Delta y$ map (right) of the optimized metasurface. (b) The phase maps under $+\hat{x}$ guided wave excitation (left) and $-\hat{x}$ guided wave excitation (right) for wavelength $\lambda = 650$ nm.

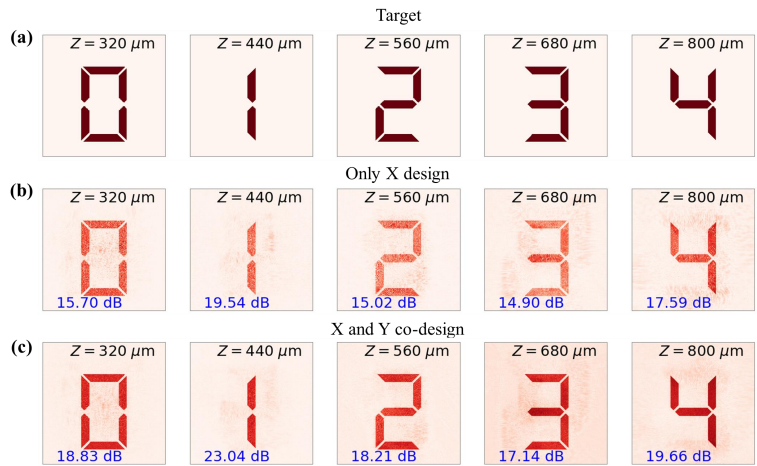

**Figure S8.** Holograms by co-design for 5 target planes. (a) The target images of digital number. (c) The results solely using $\Delta x$ as the design parameter. (d) The results using both $\Delta x$ and $\Delta y$ as the design space parameters.

# Supplementary Note 6: Assessment of the impact of fabrication error on the metasurface performance

To emulate the effects of fabrication errors, we incorporated random perturbations into the meta-atom positions. In Figure S9, panels (a) and (b) correspond to the scenarios of positional perturbations of ±10 nm and ±20 nm, respectively. Specifically, we introduced random positional perturbations following a Gaussian distribution within the specified ranges to the optimized metasurface and computed the corresponding holograms. Visual inspection of the holograms under these perturbations, as compared to Figure 3(d), reveals that the quality does not experience a substantial decline. Furthermore, we conducted a comparison of the on-chip metasurface scattering phase before and after the introduction of positional perturbations, as depicted in Figure S9(c). The findings suggest that although positional perturbations induce some phase alterations, these changes are minimal and do not significantly disrupt the integrity of the original phase gradient.

Finally, we have introduced a fluctuation of up to 10% in the position of each meta-atom and conducted model predictions. The outcomes are displayed in Figure S10. These perturbations were set at 5%, 10%, 15%, and 20% of the single pixel size, corresponding to perturbation ranges of ±20 nm, ±40 nm, ±60 nm, and ±80 nm, respectively. Figure S10(a) showcases a 10% level of random positional perturbation. Figure S10(b) elucidates the relationship between phase change and positional variations. Among the channels, blue channel experiences the most pronounced phase alteration under these positional disturbances. As depicted in Figure S10(c), as the perturbation level increases, the background noise in the holograms increases progressively, while the Peak Signal-to-Noise Ratio (PSNR) of the holograms declines. Specifically, at a 10% perturbation level, the PSNR of the holograms drops by approximately 1 dB. This observation underscores the metasurface's inherent robustness and its ability to accommodate a certain degree of fabrication error without significant performance degradation.

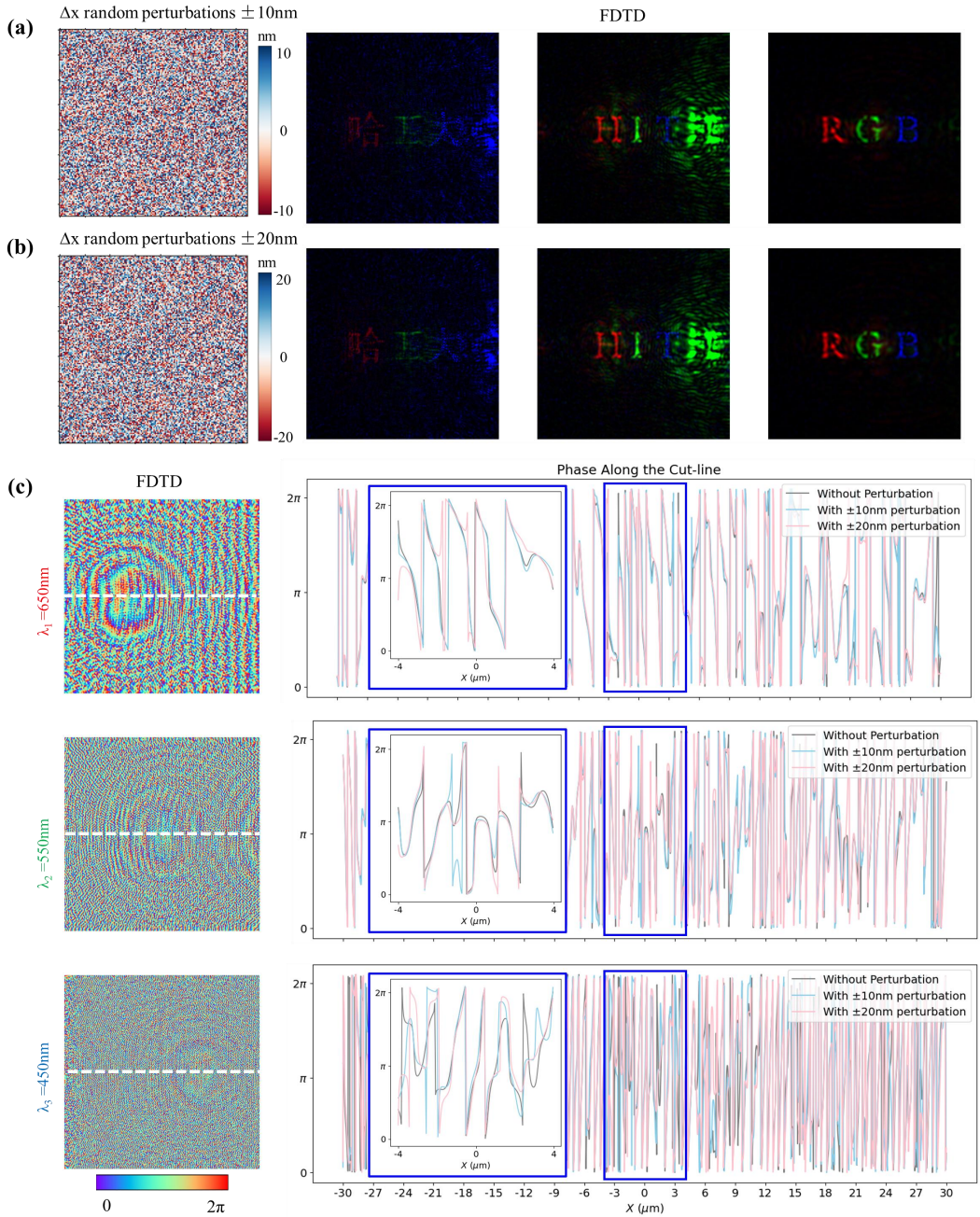

**Figure S9.** Positional perturbations on the hologram quality evaluated by FDTD simulation, in comparison to Figure 3(d). The random positional perturbations added to the metaatoms are within the range of (a) ±10 nm and (b) ±20 nm, and their corresponding holograms are also shown. (c) Left column: The phase map of Δx ±20 nm perturbation. The results are from the FDTD simulation for λ = 650 nm, λ = 550 nm, and λ = 450 nm, respectively. Right column: The phase without perturbations added, ±10 nm perturbations added, and ±20 nm perturbations added along the white dashed cutting line in (c).

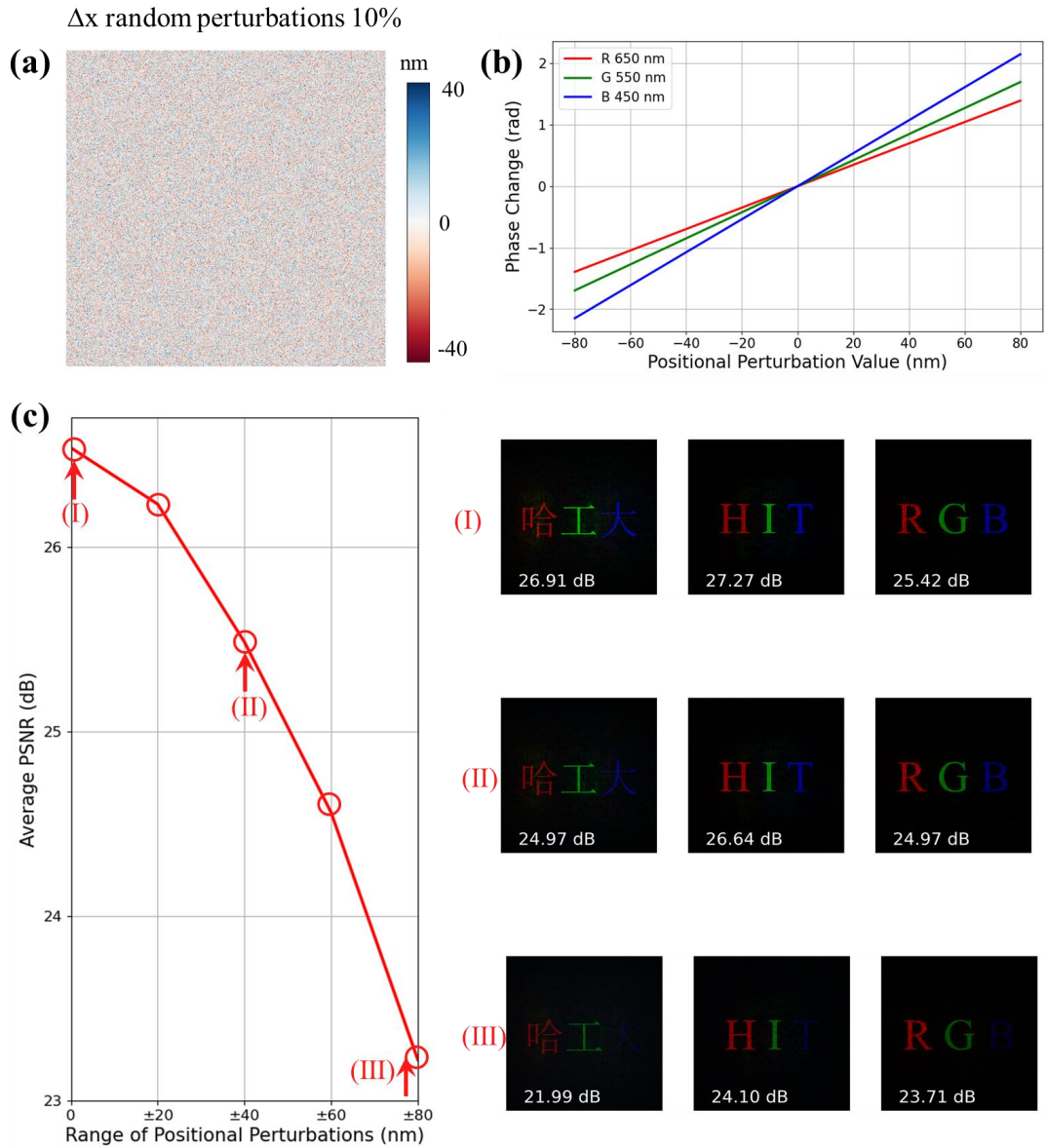

**Figure S10.** Positional perturbations on the hologram quality evaluated by the theoretical model. (a) A typical 10% level of random positional perturbation. (b) Correlation between propagation phase change and position perturbation. (c) Relationship between average PSNR of the holograms and the position perturbation.